\documentclass[12pt,authoryear,3p,times]{elsarticle}

\usepackage{amssymb}
\usepackage{amsmath}
\usepackage{mathdots}

\usepackage{relsize}
\usepackage{booktabs}
\usepackage{hyperref}
\usepackage[utf8]{inputenc}
\usepackage[T1]{fontenc} 
\usepackage{multirow}
\usepackage[table,xcdraw]{xcolor}
\usepackage[dvipsnames]{xcolor} 
\usepackage{placeins}
\usepackage{orcidlink}
\usepackage{array}
\usepackage{blkarray}
\usepackage{bigstrut}
\usepackage{multicol}

\definecolor{dred}{rgb}{0.75,0,0}

\definecolor{gray}{rgb}{0.5,0.5,0.5}

\DeclareMathOperator*{\argmax}{argmax}

\journal{Energy Economics}

\begin{document}

\begin{frontmatter}

\renewcommand{\thefootnote}{$\bigstar$}

\title{Stealing profits: Spread-based temporal hierarchy forecasting for day-ahead electricity markets}

\author[ONAS]{Arkadiusz Lipiecki \orcidlink{0000-0003-1118-0388}}
\author[Indicio]{Nikolaos Kourentzes \orcidlink{0000-0003-0211-5218}}
\author[ORBI]{Rafał Weron \orcidlink{0000-0003-1619-5239}  \corref{cor1}}
\address[ONAS]{Department of Computational Social Science, Wrocław University of Science and Technology, Poland}
\address[Indicio]{Indicio Technologies AB, Stockholm, Sweden}
\address[ORBI]{Department of Operations Research and Business Intelligence, Wrocław University of Science and Technology, Poland}
\cortext[cor1]{Corresponding author; \textit{email:} rafal.weron@pwr.edu.pl}

\begin{abstract}
Day-ahead electricity price forecasts support trading and storage decisions, but for battery arbitrage predicting intraday price spreads is more relevant than predicting individual hourly prices. Here we show that a temporal hierarchy forecasting (THieF) framework that jointly reconciles forecasts of hourly electricity prices and all intraday price spreads consistently improves performance across two major European electricity markets and three different forecasting architectures. Using five years of out-of-sample data from Germany and Spain, we obtain accuracy improvements of up to 19.7\% and profit gains of up to 10.4\% relative to unreconciled hourly price forecasts. The gains persist even for a highly accurate pretrained TabPFN foundation model. Our results demonstrate that exploiting coherent relationships between economically relevant forecasting targets can improve both predictive accuracy and decision value, and that better statistical forecasts do not necessarily imply better economic decisions.
\end{abstract}

\begin{keyword}
electricity price forecasting \sep price spread \sep temporal hierarchy forecasting (THieF) \sep forecast reconciliation \sep BESS arbitrage \sep opportunity cost

\end{keyword}

\end{frontmatter}

\begin{multicols}{2}
{\scriptsize
\tableofcontents
}
\end{multicols}

\newpage

\section{Introduction}

Day-ahead electricity price forecasts support a wide range of operational and market decisions, ranging from bidding and scheduling \citep{bel:etal:22,nar:zie:22:bidding} to trading and risk management \citep{jan:woj:22,aga:hae:kop:koz:pet:25}.
For a battery energy storage system (BESS) operator, however, the value of a forecast does not depend solely on the accuracy of the predicted price at a particular hour. What ultimately determines the profitability is the difference between the selling and buying prices, adjusted for battery efficiency and operational costs \citep{mer:oli:dej:23,lin:zhu:wid:24}. 

Consequently, the information relevant for the decision is not limited to individual hourly prices but also includes the price differences between potential buying and selling hours \citep{ser:wer:25}. This distinction is important because an accurate set of hourly price forecasts does not automatically imply accurate spread forecasts. Forecast errors for two hours may reinforce or cancel each other, and the relative ranking of the predicted prices may matter more for trading than their individual levels \citep{mac:lip:uni:26}. 

At the same time, hourly prices and price spreads are not unrelated forecasting targets. Every spread is a linear combination of two hourly prices, while the entire collection of spreads is generated from the same 24-dimensional daily price vector. Forecasts produced independently for hourly prices and spreads will generally violate these aggregation constraints. However, forecast reconciliation offers a systematic way of resolving this inconsistency. The idea is to combine independently generated base forecasts subject to a set of linear aggregation constraints, thereby producing coherent forecasts that exploit information available at different levels of a hierarchy \citep{ath:hyn:kou:pet:17,wic:etal:2019}. 

\cite{lip:bil:kou:wer:26} introduced \textit{temporal hierarchy forecasting} (THieF) for day-ahead \textit{electricity price forecasting} (EPF) by jointly modeling hourly prices and non-overlapping price blocks ranging from 2 to 24 hours. Reconciling forecasts across these temporal aggregation levels significantly improved accuracy for the German and Spanish power markets, across linear regression, shallow neural networks, gradient-boosted decision trees, and a Mitra foundation model. The gains from \textit{Block THieF} reached 5.5\% for hourly prices and 13.4\% for baseload prices. The intuition behind these improvements is that the method is able to reduce modeling uncertainty by reconciling base forecasts from multiple temporal aggregation levels, and, taking advantage of the temporal aggregation, enable forecasting models to extract more structural details of the electricity price series. 

Rather than adding another temporal aggregation level, the present study introduces a fundamentally different source of information into the reconciliation framework -- direct forecasts of intraday price spreads. \textit{Spread THieF} combines forecasts for the 24 hourly prices with forecasts for the 276 spreads between potential buying and selling hours. Reconciliation blends these two sets of forecasts while enforcing coherence. In this sense, the approach goes beyond simply computing spreads from reconciled hourly prices: it allows the directly estimated spread models to contribute information that may be difficult to recover from price-level forecasts alone. We demonstrate that the proposed modification leads to a new type of information combination that was impossible with standard forecast combination and Block THieF, specifically capturing the relative evolution of prices within the day. 
Standard forecasting models each hourly price directly. Block THieF enriches these forecasts with information obtained from multiple temporal aggregations, thereby reducing modelling uncertainty. Spread THieF introduces a different source of information: direct forecasts of all intraday price differences. Reconciliation then ensures that the resulting hourly price forecasts are jointly coherent with these relative intraday price movements.

We evaluate the proposed approach using day-ahead electricity prices from the German EPEX-DE and Spanish OMIE markets over the 2021-2025 test period. To assess whether the results depend on model architecture, we consider three forecasting methods of increasing complexity: a linear autoregressive model with exogenous variables (ARX), a shallow nonlinear autoregressive neural network (NARX), and the pretrained TabPFN-2 foundation model. This allows us to evaluate whether the benefits of Spread THieF persist across substantially different forecasting architectures and levels of baseline accuracy.


For each architecture, we compare four approaches: unreconciled hourly price forecasts, unreconciled spread forecasts \cite[as in][]{ser:wer:25}, Block THieF forecasts \citep{lip:bil:kou:wer:26}, and the proposed Spread THieF forecasts. Predictive performance is evaluated using the mean absolute error (MAE) and root mean squared error (RMSE), with differences in predictive ability formally assessed using conditional predictive ability (CPA) tests \citep{gia:whi:06}. Economic performance is evaluated through realized profits from a BESS-arbitrage strategy similar to those in \cite{mar:nar:wer:zie:23} and \cite{uni:25}, as well as through opportunity costs relative to a perfect-foresight Oracle benchmark.

Our contribution is threefold. First, we formulate price spread prediction as a coherent hierarchical forecasting problem linking hourly prices with all feasible intraday spreads. Second, we introduce a new combination mechanism, building on minimum-trace forecast reconciliation, that incorporates direct forecasts of intraday price differences and captures information on the relative evolution of prices within the day. Third, we demonstrate that the resulting forecasts are superior not only in terms of standard statistical error metrics but also in terms of their economic value for BESS trading.

The remainder of the paper is structured as follows. Section~\ref{sec:THieF} introduces the THieF framework, first recalling the conventional block hierarchy and then extending it to incorporate price spread information; it also describes the forecast reconciliation procedure and covariance matrix estimation. Section~\ref{sec:Datasets} presents the German and Spanish electricity market datasets and the rolling window forecasting setup. Section~\ref{sec:base:forecasts} describes the three forecasting approaches used to generate hourly price, block price, and price spread forecasts: ARX, NARX, and TabPFN. Section~\ref{sec:Eval} introduces the statistical and economic evaluation measures, including the BESS-arbitrage strategy and opportunity cost measure. Section~\ref{sec:results} presents the empirical results, and Section~\ref{sec:conclusions} concludes.

\section{Temporal hierarchies and reconciliation}
\label{sec:THieF}

The main idea behind temporal hierarchy forecasting (THieF) is to (i)~construct a temporal hierarchy to help identify the structure, (ii)~model the different aggregate views separately, and (iii)~reconcile the modelled information into the final prediction \citep{ath:hyn:kou:pet:17}. 
The key advantages of THieF are that it mitigates modelling uncertainty by combining forecasts across multiple temporal aggregation levels and leverages the filtering effect of temporal aggregation to better estimate components that may be difficult to capture at a single frequency \citep{kou:pet:tra:14}. \cite{kou:ros:bar:17} showed that using multiple temporal aggregation levels performs better than using any single aggregation. This is partly explained by the accuracy benefits expected from combining forecasts \citep{wan:hyn:li:kan:23}. However, a challenge with forecast combination is the appropriate construction of the pool of alternative forecasts and the estimation of the combination weights. THieF simplifies this substantially, by benefiting from the highly structured hierarchy.

\subsection{Construction of the block hierarchy}
\label{ssec:BlockThieF:S}

\begin{figure}[tbp]
\centering
\includegraphics[width=\linewidth]{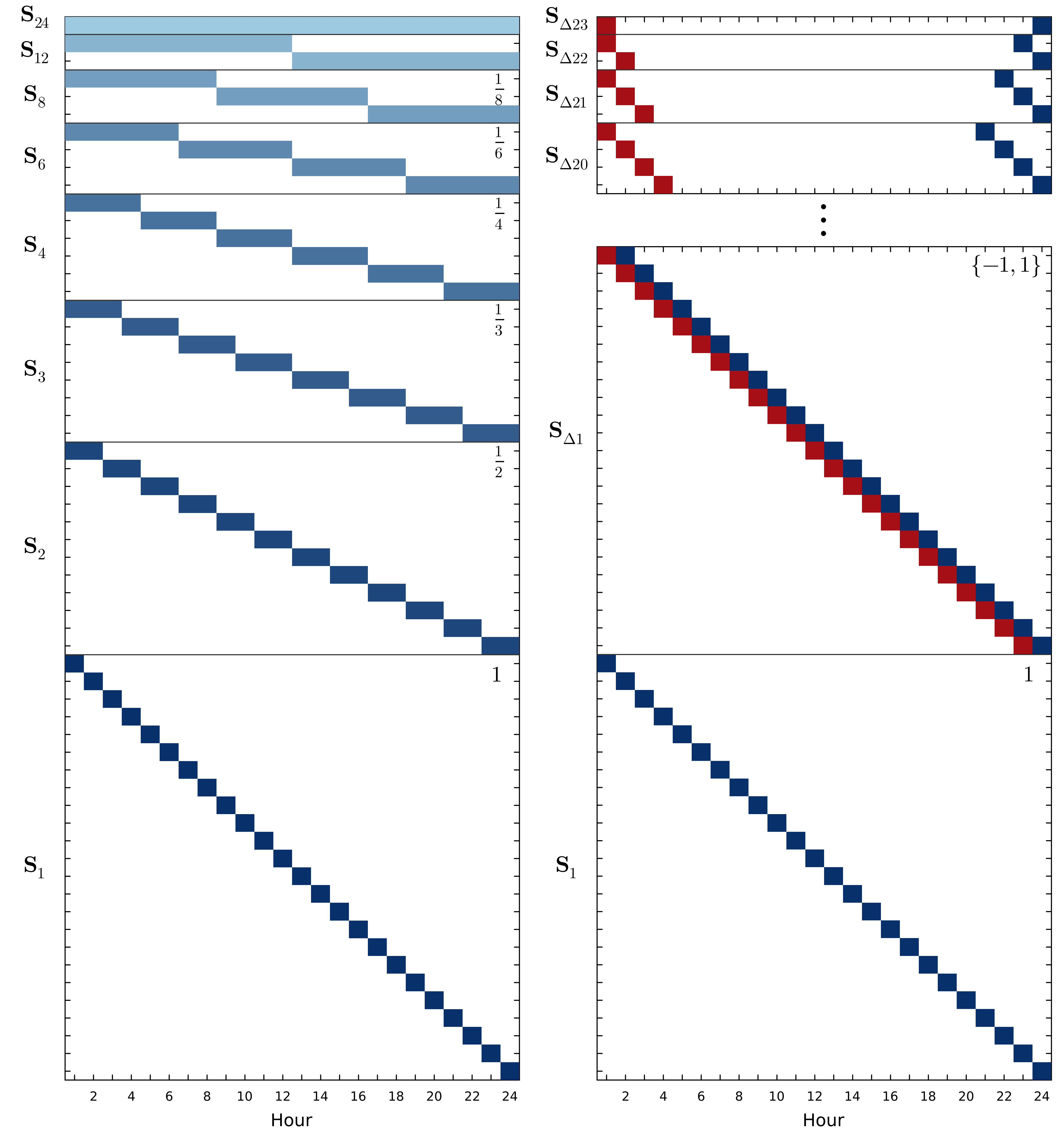}
\caption{The summing matrices $\mathbf{S}_{block}$ (\textit{left panel}) and $\mathbf{S}_{spread}$ (\textit{right panel}) defined in Eq.~\eqref{eqn:blockSmatrix} and Eq.~\eqref{eqn:diffSmatrix}, respectively. Blue blocks denote positive elements with shading indicating weights ranging from $\frac{1}{24}$ for $\mathbf{S}_{24}$ to $1$ for $\mathbf{S}_1$, while red blocks denote negative values equal to $-1$. The weights occurring in each submatrix are annotated in its top-right corner.}
\label{fig:summing_matrix}
\end{figure}

Before we introduce the spread hierarchy, let us briefly recall after \cite{lip:bil:kou:wer:26} how the block hierarchy is constructed. Let $y_t$ be an observation of a time series at period $t$, of frequency $m$. Let $k = \{k_i\}_{i=1}^p$ be the factors of $m$, ordered from the largest to the smallest. For example, for hourly data $m=24$ and $k_i \in \{24, 12, 8, 6, 4, 3, 2, 1\}$ and $p=8$. Non-overlapping temporal aggregates at level $k$ are constructed using mean-aggregation:
\begin{equation}
y_{j}^{[k]} = \frac{1}{k} \sum_{t=t^*}^{t^*+k-1} y_{t}, \quad \text{where } t^* = (j-1)k + 1,
\end{equation}
\noindent where $t^*$ denotes the starting time index of the $j$-th aggregation block at temporal scale $k$, where $j = 1, \dots, \frac{m}{k}$ and ensures that there are always complete blocks of aggregation. The full hierarchy of observations at all temporal aggregation levels can be represented through a structural linear relationship:
$\mathbf{y} = \mathbf{S} \mathbf{b}$, where $\mathbf{b}$ is the $m \times 1$ vector of observations at the most disaggregated level (hourly), $\mathbf{y}$ is the $K \times 1$ stacked vector of all temporal levels, $K = \sum_{i=1}^p \frac{m}{k_i}$, 
and $\mathbf{S}$ is the $K \times m$ \emph{summing matrix}:
\begin{equation}
\mathbf{S} = \begin{bmatrix} \mathbf{S}_{k_p}^\top & \mathbf{S}_{k_{p-1}}^\top & \dots & \mathbf{S}_{k_1}^\top \end{bmatrix}^\top,
\end{equation}
where $\mathbf{S}^\top$ is the transpose of $\mathbf{S}$. Each block matrix $\mathbf{S}_k$ maps the disaggregated $m$ values to their mean temporal aggregate at frequency $k$:
\begin{equation}
\mathbf{S}_k = f_k \left( \mathbf{I}_{m/k} \otimes \mathbf{J}_{1, k} \right),
\end{equation}
where $f_k = 1/k$ is the mean-scaling factor, $\mathbf{I}_{m/k}$ is the identity matrix of dimension $m/k$, $\mathbf{J}_{1, k}$ is a $1 \times k$ row vector of ones, and $\otimes$ denotes the Kronecker product. Specifically for hourly data with $m=24$ we have:
\begin{equation}
\mathbf{S}_{block} = \begin{bmatrix} \mathbf{S}_{24}^\top & \mathbf{S}_{12}^\top & \mathbf{S}_{8}^\top & \mathbf{S}_{6}^\top & \mathbf{S}_{4}^\top & \mathbf{S}_{3}^\top & \mathbf{S}_{2}^\top & \mathbf{S}_{1}^\top  \end{bmatrix}^\top,
\label{eqn:blockSmatrix}
\end{equation}
where the first three submatrices are:
\setcounter{MaxMatrixCols}{24}
\setlength{\arraycolsep}{2pt}
\begin{align}
\mathbf{S}_{24} = \tfrac{1}{24} &
    \begin{bmatrix} 
        1 & 1 & 1 & 1 & 1 & 1 & 1 & 1 & 1 & 1 & 1 & 1 & 1 & 1 & 1 & 1 & 1 & 1 & 1 & 1 & 1 & 1 & 1 & 1
    \end{bmatrix}, \\
\mathbf{S}_{12} = \tfrac{1}{12} &
    \begin{bmatrix}
        1 & 1 & 1 & 1 & 1 & 1 & 1 & 1 & 1 & 1 & 1 & 1 & 0 & 0 & 0 & 0 & 0 & 0 & 0 & 0 & 0 & 0 & 0 & 0 \\
        0 & 0 & 0 & 0 & 0 & 0 & 0 & 0 & 0 & 0 & 0 & 0 & 1 & 1 & 1 & 1 & 1 & 1 & 1 & 1 & 1 & 1 & 1 & 1
    \end{bmatrix}, \\
\mathbf{S}_{8} = \tfrac{1}{8} &
    \begin{bmatrix}
        1 & 1 & 1 & 1 & 1 & 1 & 1 & 1 & 0 & 0 & 0 & 0 & 0 & 0 & 0 & 0 & 0 & 0 & 0 & 0 & 0 & 0 & 0 & 0 \\
        0 & 0 & 0 & 0 & 0 & 0 & 0 & 0 & 1 & 1 & 1 & 1 & 1 & 1 & 1 & 1 & 0 & 0 & 0 & 0 & 0 & 0 & 0 & 0 \\
        0 & 0 & 0 & 0 & 0 & 0 & 0 & 0 & 0 & 0 & 0 & 0 & 0 & 0 & 0 & 0 & 1 & 1 & 1 & 1 & 1 & 1 & 1 & 1
    \end{bmatrix}, 
\end{align}
and the last one is simply the identity matrix $\mathbf{S}_{1} = \mathbf{I}_{24}$. The full summing matrix for Block THieF is visualized in the left panel of Figure~\ref{fig:summing_matrix}.

\subsection{Including spread information in the hierarchy}
\label{ssec:SpreadThieF:S}

The spread-augmented summing matrix maps the $m \times 1$ vector of hourly prices $\mathbf{b}$ to the stacked vector containing all pairwise spreads and hourly prices.
The matrix has dimensions $\left(\binom{m}{2} + m\right) \times m$ and can be partitioned vertically into two main parts: the spread matrix $\mathbf{S}_{\Delta}$ and the identity matrix $\mathbf{I}_m$. Each row in $\mathbf{S}_{\Delta}$ corresponds to a specific pair of periods $(i, j)$ with $1 \le i < j \le m$, representing a spread between period $i$ and period $j$: $\Delta y_{i,j} = y_j - y_i$. The total number of spreads is $\binom{m}{2} = \frac{m(m-1)}{2}$. We can group them by lag length $l = j - i = m-1, m-2, \dots, 1$, to obtain a decomposition of $\mathbf{S}_{\Delta}$ in the form of  $\begin{bmatrix}\mathbf{S}_{\Delta(m-1)}^\top\;...\; \mathbf{S}_{\Delta 1}^\top\end{bmatrix}^\top$, where the elements of $\mathbf{S}_{\Delta l}$ are given by:
\begin{eqnarray}
    (\mathbf{S}_{\Delta l})_{i,j} = \begin{cases} -1 & \text{if } j=i, \\  
    1 & \text{if } j=i+l, \\  
    0 & \text{otherwise}.  \end{cases}
\end{eqnarray}
For hourly electricity market data with $m = 24$, there are $\binom{24}{2} = 276$ spread combinations. The overall dimension of the summing matrix is $300 \times 24$ and includes 276 spread rows and 24 price-level rows:
\begin{equation}
\mathbf{S}_{spread} = 
\begin{bmatrix}  \mathbf{S}_{\Delta 23}^\top & \mathbf{S}_{\Delta 22}^\top & \dots & \mathbf{S}_{\Delta 1}^\top &  \mathbf{S}_{1}^\top  \end{bmatrix}^\top,
\label{eqn:diffSmatrix}
\end{equation}
where the submatrices for 23-hour, 22-hour and 1-hour spreads are respectively given by:
\setcounter{MaxMatrixCols}{24}
\newcolumntype{C}{>{\raggedleft\arraybackslash$}p{0.9em}<{$}}
\begin{align}
\mathbf{S}_{\Delta 23} & = 
    \left[ \begin{array}{*{24}{C}} 
        -1 & 0 & 0 & 0 & 0 & 0 & 0 & 0 & 0 & 0 & 0 & 0 & 0 & 0 & 0 & 0 & 0 & 0 & 0 & 0 & 0 & 0 & 0 & 1 
    \end{array} \right], \\
\mathbf{S}_{\Delta 22} & = 
    \left[ \begin{array}{*{24}{C}}
        -1 & 0 & 0 & 0 & 0 & 0 & 0 & 0 & 0 & 0 & 0 & 0 & 0 & 0 & 0 & 0 & 0 & 0 & 0 & 0 & 0 & 0 & 1 & 0 \\
        0 & -1 & 0 & 0 & 0 & 0 & 0 & 0 & 0 & 0 & 0 & 0 & 0 & 0 & 0 & 0 & 0 & 0 & 0 & 0 & 0 & 0 & 0 & 1 
    \end{array} \right], \\
\mathbf{S}_{\Delta 1} & = 
    \left[ \begin{array}{*{24}{C}}     
        -1 & 1 & 0 & 0 & 0 & 0 & 0 & 0 & 0 & 0 & 0 & 0 & 0 & 0 & 0 & 0 & 0 & 0 & 0 & 0 & 0 & 0 & 0 & 0 \\
        0 & -1 & 1 & 0 & 0 & 0 & 0 & 0 & 0 & 0 & 0 & 0 & 0 & 0 & 0 & 0 & 0 & 0 & 0 & 0 & 0 & 0 & 0 & 0 \\
        \multicolumn{24}{c}{\ddots} \\
        0 & 0 & 0 & 0 & 0 & 0 & 0 & 0 & 0 & 0 & 0 & 0 & 0 & 0 & 0 & 0 & 0 & 0 & 0 & 0 & 0 & -1 & 1 & 0 \\
        0 & 0 & 0 & 0 & 0 & 0 & 0 & 0 & 0 & 0 & 0 & 0 & 0 & 0 & 0 & 0 & 0 & 0 & 0 & 0 & 0 & 0 & -1 & 1
    \end{array} \right].
\end{align}
Like in Eq.~\eqref{eqn:blockSmatrix}, the last part is simply the identity matrix $\mathbf{S}_{1} = \mathbf{I}_{24}$. The complete summing matrix for Spread THieF is illustrated in the right panel of Figure~\ref{fig:summing_matrix}.

\subsection{Forecast reconciliation}
\label{ssec:reconciliation}

Base forecasts are constructed for each series in the hierarchy, which are then stacked: $\hat{\mathbf{y}} = \begin{bmatrix} \hat{\mathbf{y}}_{k_p}^\top & \hat{\mathbf{y}}_{k_{p-1}}^\top & \dots & \hat{\mathbf{y}}_{k_1}^\top \end{bmatrix}^\top$ for Block THieF and $\hat{\mathbf{y}} = \begin{bmatrix} \hat{\mathbf{y}}_{\Delta(m-1)}^\top & \hat{\mathbf{y}}_{\Delta(m-2)}^\top & \dots & \hat{\mathbf{y}}_{\Delta 1}^\top  & \hat{\mathbf{y}}_{k_1}^\top \end{bmatrix}^\top$ for Spread THieF, where $\hat{\mathbf{y}}_{\Delta l}$ is the vector of forecasts for spreads between load periods $l$ hours apart. Since the individual series are forecast independently, the resulting base forecasts need not satisfy the linear relationships encoded by $\mathbf{S}$, i.e., $\hat{\mathbf{y}} \neq \mathbf{S}\hat{\mathbf{b}}$. This means that aggregating the bottom-level forecasts may differ from forecasting the aggregate series directly, or that a directly forecast spread may differ from the difference between the corresponding independently forecast hourly prices.

Following \citep{ath:hyn:kou:pet:17}, we can take advantage of these differences to obtain better forecasts. The reconciled forecasts are given by \citep{wic:etal:2019}:
\begin{equation}
\label{eq:SGy}
\tilde{\mathbf{y}} = \mathbf{S} \mathbf{G} \hat{\mathbf{y}}, ~~ \text{where} ~~ \mathbf{G} = \left( \mathbf{S}^\top \mathbf{W}^{-1} \mathbf{S} \right)^{-1} \mathbf{S}^\top \mathbf{W}^{-1},
\end{equation}
$\mathbf{W}$ is the $K \times K$ covariance matrix of the base forecast errors, and $\mathbf{W}^{-1}$ is the inverse of $\mathbf{W}$.

\subsection{Covariance matrix estimation}
\label{ssec:Shrinkage}

Forecast reconciliation requires an estimate of the covariance matrix of the base forecast errors. Estimating this matrix directly can be challenging, particularly for Spread THieF, where $\mathbf{W}$ has dimension $300\times300$. Moreover, the forecast errors of hourly prices and price spreads may differ substantially in scale and exhibit strong cross-correlations. To mitigate estimation issues, we use highly restricted approximations, or shrinkage estimators \citep{pri:sve:kou:21}.

Our main specification is the diagonal-target shrinkage estimator of \cite{sch:str:05}, which shrinks the sample covariance matrix towards its diagonal:
\begin{equation}
    \widehat{\mathbf{W}}_{\mathrm{shr}} = (1-\lambda)\widehat{\mathbf{W}} +    \lambda\,\mathrm{diag}(\widehat{\mathbf{W}}),
\label{eqn:shrinkage}
\end{equation}
where $\lambda\in[0,1]$ is the shrinkage intensity. This approach preserves the estimated error variances while shrinking the off-diagonal covariances towards zero, thereby reducing estimation uncertainty without ignoring differences in scale across the series. For $\lambda=1$, Eq.~\eqref{eqn:shrinkage} reduces to variance scaling, corresponding to a weighted least-squares reconciliation that ignores cross-series error dependence. The sample covariance matrix $\mathbf{W}$, as well as the optimal shrinkage $\lambda$ are re-estimated daily based on the in-sample forecast errors from the entire training window, i.e., 1085 most recent days, see~\citep{sch:str:05} for details on deriving the optimal shrinkage intensity. As a robustness check, we also considered the constant-correlation shrinkage estimator of \citet{led:wol:04}. It yields qualitatively similar results, although the diagonal-target estimator generally performs slightly better. 

\begin{figure*}
    \centering
    \includegraphics[width=\linewidth]{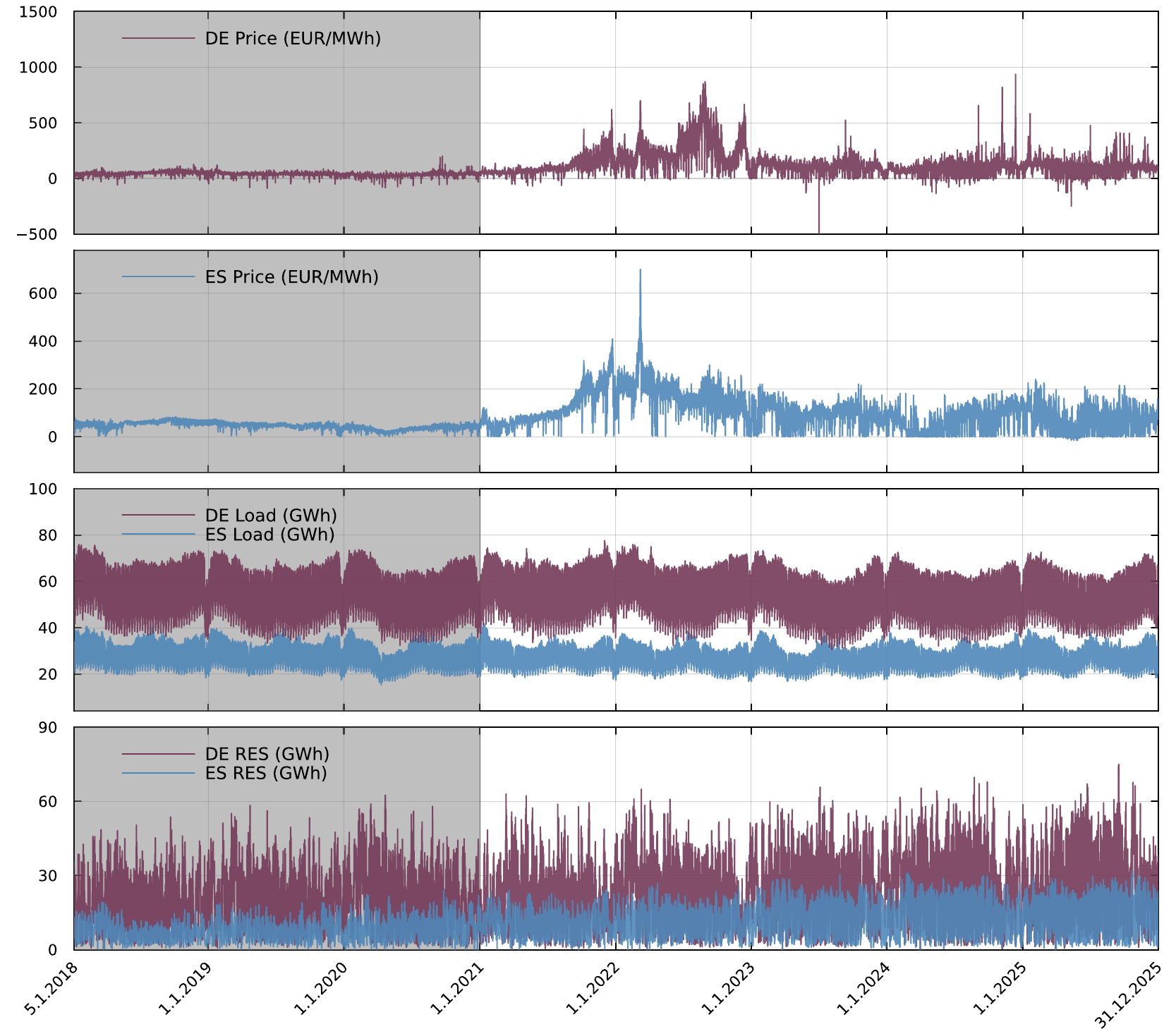}
   \caption{\textit{Top two panels:} German (DE) and Spanish (ES) day-ahead electricity prices from 5 January 2018 to 31 December 2025. \textit{Middle panels:} Day-ahead load forecasts in both countries. \textit{Bottom panel:} Day-ahead renewable generation forecasts in both countries, including total wind generation and solar generation. The shaded area indicates the initial training window used for generating forecasts for the first day of the test period, 1 January 2021.}
    \label{fig:datasets}
\end{figure*}

\section{Datasets}
\label{sec:Datasets}

To ensure a sound assessment of the THieF approach, we consider two major European power markets: EPEX-DE (Germany) and OMIE (Spain). Both datasets span eight years (05.01.2018-31.12.2025, see Figure \ref{fig:datasets}) and include publicly available data downloaded in July 2026:
\begin{itemize}
    \item Three series specific to each market (hourly resolution) -- day-ahead prices $p_{d,h}$, day-ahead load $\hat{L}_{d, h}$ and renewable generation forecasts $\hat{R}_{d,h}$. We define the latter as the sum of onshore wind, offshore wind, and solar generation (source: ENTSO-E Transparency Platform, \url{https://transparency.entsoe.eu}). 
    
    \item Two series common to both markets (daily resolution) -- the last known (hence the $d-2$ time index) closing prices of nearest-to-delivery futures contracts written on natural gas from the Title Transfer Facility in the Netherlands, $\text{TTF}_{d-2}$, and API2 coal, $\text{API}_{d-2}$ (source: Investing.com, \url{https://www.investing.com/}). 
\end{itemize}
Note that, compared to \cite{ser:wer:25} and \cite{lip:bil:kou:wer:26}, the datasets are extended by one year and include solar generation forecasts in $\hat{R}_{d,h}$. Also note that due to revisions of past data on the ENTSO-E Transparency Platform, the data used in this study differs from that considered in \cite{ser:wer:25}, \cite{lip:bil:kou:wer:26}, and \cite{mac:lip:uni:26}, particularly the German day-ahead prices in 2024; the latter three articles used data downloaded from ENTSO-E in January 2025.

The first 1092 days (until 31.12.2020) are the initial training window. After accounting for the lags in prices, see Eq.~\eqref{eqn:forecast} below, this corresponds to $1092-7=1085$ data points for training. Each day the training window is rolled forward by 24 hours.
The remaining five-year period (starting 01.01.2021) is a challenging test set that covers the later stages of the COVID-19 pandemic, the 2021-2022 energy crisis, and the Russian invasion of Ukraine. This period is also marked by the appearance of negative price spikes in Germany. For instance, on Sunday 02.07.2023 at hour 15 the price dropped to $-$500 EUR/MWh due to low demand and high renewable generation. On the other hand, negative prices were not seen in the Spanish market before 2024. In 2024 the prices dropped only a few EUR below zero, but in 2025 the lowest observed price was $-$15 EUR.

\section{Forecasting models}
\label{sec:base:forecasts}

To demonstrate the versatility of our approach, we generate base forecasts using three models with distinct architectures: a linear regression, a shallow neural network, and a pretrained transformer-based tabular foundation model. The first two are commonly used as benchmarks in EPF \citep{zie:wer:18,mac:nit:wer:21,bil:gia:del:rav:23,uni:mac:23,ghe:zie:25}. The third is a recent and highly competitive model for EPF tasks \citep{lip:wer:26:FM:arXiv}.

All models use a similar set of features and are estimated independently for each series using a 3-year rolling window of past observations, which is rolled forward by one day after each forecast. Overall, we consider 336 series: 24 hourly prices, 36 block prices (12 $\times$ 2H, 8 $\times$ 3H, 6 $\times$ 4H, 4 $\times$ 6H, 3 $\times$ 8H, 2 $\times$ 12H, and 1 $\times$ 24H), and 276 price spreads.

For the 24 hourly and 36 block price series, the forecast $\hat{p}_{d,h}$ for day $d$ and series $h$ is obtained as a function of 14 features:
\begin{align}
    \hat{p}_{d,h}
    = f\Big(p_{d-1,h}, \ldots, p_{d-7,h}, ~~ p_{d-1}^{min}, p_{d-1}^{max}, \widehat{L}_{d,h}, \widehat{R}_{d,h},     \text{API2}_{d-2}, \text{TTF}_{d-2}, \text{DoW}_d \Big)
\label{eqn:forecast}
\end{align}
where
\begin{itemize}
    \item $p_{d-i, h}$ are the lagged prices of the same series $h$ over the previous seven days, $i=1,\ldots,7$,  
    \item $p_{d-1}^{min}$ and $p_{d-1}^{max}$ are the minimum and maximum hourly prices on the previous day, 
    \item $\widehat{L}_{d, h}$ and $\widehat{R}_{d, h}$ are the day-ahead load and renewable generation forecasts corresponding to series $h$, 
    \item $\text{API}_{d-2}$ and $\text{TTF}_{d-2}$ are the last known closing prices, observed on day $d-2$, of the nearest-to-delivery monthly API2 coal and yearly TTF natural gas futures contracts, respectively, and 
    \item $\text{DoW}_d$ is a categorical variable representing the day of the week. 
\end{itemize}
For the 276 price spread series, the forecast $\widehat{\Delta p}_{d,h_1,h_2}$ for day $d$ and a pair of hours $h_1<h_2$ is likewise obtained from 14 features, with the lagged day-ahead prices and the load and renewable generation forecasts replaced by the corresponding spreads:
\begin{align}
    \widehat{\Delta p}_{d,h_1,h_2} = & ~~ f\Big(
    \Delta p_{d-1,h_1,h_2}, \ldots, \Delta p_{d-7,h_1,h_2},
    p_{d-1}^{\min}, p_{d-1}^{\max}, \nonumber \\ 
    & ~~~~~~ \widehat{\Delta L}_{d,h_1,h_2},
    \widehat{\Delta R}_{d,h_1,h_2},
    \text{API2}_{d-2}, \text{TTF}_{d-2},
    \text{DoW}_d \Big),
\label{eqn:forecast:spread}
\end{align}
where
\begin{itemize}
    \item $\Delta p_{d-i,h_1,h_2}$ are the lagged price spreads for the same pair of hours $(h_1,h_2)$ over the previous seven days, $i=1,\ldots,7$, and
    \item $\widehat{\Delta L}_{d,h_1,h_2}$ and $\widehat{\Delta R}_{d,h_1,h_2}$ are the corresponding differences in the day-ahead load and renewable generation forecasts between hours $h_1$ and $h_2$.
\end{itemize}

\subsection{ARX}
\label{ssec:ARX}

The first and simplest forecasting approach considered in our study is an AutoRegression with eXogenous variables (ARX), estimated separately for each series using ordinary least squares. Separate ARX models are fitted for the 24 hourly prices, the 36 block prices, and the 276 price spreads.
For the former 60 series, the ARX specification is
\begin{align}
    \hat{p}_{d,h} ={}& \beta_1 p_{d-1,h} + \cdots + \beta_7 p_{d-7,h} + \beta_8 p_{d-1}^{\min} + \beta_9 p_{d-1}^{\max}  + \beta_{10}\widehat{L}_{d,h} + \beta_{11}\widehat{R}_{d,h}  \nonumber\\
    & + \beta_{12}\text{API2}_{d-2} + \beta_{13}\text{TTF}_{d-2} + \beta_{14}D_d^{(1)} + \cdots + \beta_{20}D_d^{(7)},
\label{eqn:ARX:price}
\end{align}
with the $\text{DoW}_d$ categorical variable transformed using one-hot encoding into seven binary variables $D_d^{(1)}, \ldots, D_d^{(7)}$ corresponding to the days of the week. 
For the price spread series, a separate ARX model is estimated for each pair of hours $h_1<h_2$:
\begin{align}
    \widehat{\Delta p}_{d,h_1,h_2} ={}& \beta_1 \Delta p_{d-1,h_1,h_2} + \cdots + \beta_7 \Delta p_{d-7,h_1,h_2}   + \beta_8 p_{d-1}^{\min} + \beta_9 p_{d-1}^{\max} + \beta_{10}\widehat{\Delta L}_{d,h_1,h_2} + \beta_{11}\widehat{\Delta R}_{d,h_1,h_2} \nonumber\\
    &+ \beta_{12}\text{API2}_{d-2} + \beta_{13}\text{TTF}_{d-2}
    + \beta_{14}D_d^{(1)} + \cdots + \beta_{20}D_d^{(7)}.
\label{eqn:ARX:spread}
\end{align}
For more robust parameter estimation, following \cite{uni:wer:zie:18} and \cite{lag:mar:sch:wer:21}, the predictors and response are preprocessed using the inverse hyperbolic sine transformation, $\operatorname{asinh}\left(\frac{y-\hat{\mu}_y}{\hat{\sigma}_y}\right)$, where $\hat{\mu}_y$ and $\hat{\sigma}_y$ are the sample mean and standard deviation of variable $y$, respectively, estimated over the training window. The transformation is applied independently to each continuous feature and the target variable (with lags of the target transformed using the same mean and standard deviation as the target). The day-of-week categorical variable is represented by seven dummy variables.
Before forecast reconciliation and evaluation, all forecasts are transformed back to the original scale using the inverse transformation.

\subsection{NARX}
\label{ssec:NARX}

The nonlinear ARX (NARX) model uses a feedforward neural network to represent the functional relationships in Eqs.~\eqref{eqn:forecast} and \eqref{eqn:forecast:spread}. As for the ARX, separate models are estimated for the hourly price, block price, and price spread series.
Following previous EPF applications, the network contains one hidden layer with 5 neurons and hyperbolic tangent activation, and a linear output layer \citep{hub:mar:wer:19,lip:bil:kou:wer:26}. The network weights are optimized using the Levenberg-Marquardt algorithm with early stopping based on a 10\% validation set. The final forecast is obtained from a committee machine of 10 independently trained networks \citep{mar:uni:wer:19}, with the ensemble members differing in their random weight initialization and validation split. The forecast for each series is the average of the 10 individual network predictions. 
As for the ARX model, all continuous predictors and the target variable are preprocessed using the inverse hyperbolic sine transformation, applied independently to each feature, and transformed back to the original scale before reconciliation and evaluation. Similarly, the day-of-week categorical variable is represented by seven dummy variables.

\subsection{TabPFN}
\label{ssec:TabPFN}

TabPFN is a family of foundation models developed by Prior Labs for tabular classification and regression tasks~\citep{hol:etal:22}. The models are encoder-only transformers pretrained on synthetic data within the Prior-Fitted Network (PFN) learning paradigm~\citep{mul:etal:24}.
In this study, we use TabPFN-2, released in January 2025, because it is distributed under the Prior Labs license that permits commercial applications, thereby increasing the practical relevance of our study. The newer TabPFN-3 and TabPFN-TS-3 variants, released in May 2026, are not available for commercial use free of charge; see \cite{lip:wer:26:FM:arXiv} for a recent comparison of foundation models for EPF.

As for ARX and NARX, TabPFN-2 is applied independently to each hourly price, block price, and price spread series, using the feature sets defined in Eqs.~\eqref{eqn:forecast} and \eqref{eqn:forecast:spread}. Its architecture consists of a linear encoder followed by 12 TabPFN layers containing column-wise attention, row-wise attention, and a multilayer perceptron~\citep{hol:etal:25}. Forecasts are generated in the so-called \textit{zero-shot} mode, i.e., without task-specific tuning of the model weights or hyperparameters.

TabPFN-2 relies on in-context learning, with the observations from the 3-year rolling window and their corresponding predictors serving as the support set. To generate a forecast for day $d$, the predictors for the target series are passed together with this support set through the pretrained model, which produces a predictive distribution. The point forecast is taken as the expected value of this distribution. Each TabPFN-2 prediction is an ensemble of 8 estimators that differ in internal data preprocessing and feature ordering. This procedure is handled internally by the TabPFN library and mitigates the permutation asymmetry of the architecture. Unlike for ARX and NARX, we do not apply any additional transformations to the inputs or target before passing them to TabPFN-2.

\section{Evaluation metrics}
\label{sec:Eval}

\subsection{Statistical error measures}
\label{ssec:Stat:Errors}

To assess the forecasting accuracy we consider the two standard measures for evaluating point predictions, i.e., the root mean squared error (RMSE) and the mean absolute error (MAE):
\begin{eqnarray}
    \text{RMSE} = & \sqrt{\frac{1}{24T}\sum_{d=\tau+1}^{\tau+T} \sum_{h=1}^{24} (\hat{p}_{d,h} - p_{d,h})^2},\\
    \text{MAE} = & \frac{1}{24T}\sum_{d=\tau+1}^{\tau+T} \sum_{h=1}^{24} |\hat{p}_{d,h} - p_{d,h}|,
\end{eqnarray}
where $\tau$ indicates the last day in the training sample (2020.12.31) and $T=1826$ indicates the number of days in the test sample. Additionally, to evaluate the predictive power at a higher aggregation level, we report the error measures corresponding to the average daily or baseload price $\bar{p}_d$:
\begin{eqnarray}
    \text{dRMSE} = & \sqrt{\frac{1}{T}\sum_{d=\tau+1}^{\tau+T}(\hat{\bar{p}}_{d} - \bar{p}_{d})^2},\\
    \text{dMAE} = & \frac{1}{T}\sum_{d=\tau+1}^{\tau+T}|\hat{\bar{p}}_{d} - \bar{p}_{d}|.
\end{eqnarray}
To measure the improvement in forecast performance relative to the Base forecasts, we use the skill score~\citep{ras:ler:18}:
\begin{equation}
\mathrm{SS}_{M}=100\left(1-\frac{M_{\mathrm{method}}}{M_{\mathrm{Base}}}\right)\%,
\end{equation}
where $M \in \{\mathrm{RMSE},\mathrm{MAE},\mathrm{dRMSE},\mathrm{dMAE}\}$, and $M_{\mathrm{method}}$ and $M_{\mathrm{Base}}$ denote the corresponding values for the evaluated method and the Base forecasts, respectively. Positive values indicate an improvement over the Base forecasts, while negative values indicate worse performance.

Note that we do not separately evaluate errors for the 276 individual price spreads, since their relevance for trading decisions differs substantially. BESS arbitrage depends primarily on the spread associated with the selected charging and discharging hours, while for imperfect battery efficiency the economically relevant quantity is an efficiency-adjusted combination of buying and selling prices rather than the unadjusted spread.

\subsection{Testing for differences in predictive ability}
\label{ssec:Stat:CPA}

To formally compare the predictive performance of Spread THieF with that of the base forecasts and Block THieF, we apply the test of \textit{conditional predictive ability} \citep[CPA;][]{gia:whi:06}. Following \cite{lag:mar:sch:wer:21} and \cite{lip:uni:wer:24}, we use its multivariate variant that takes into account the 24-dimensional structure of day-ahead electricity price forecasts. The test aggregates the hourly loss differentials within each day and is applied separately using the daily MAE and MSE. In each case, the one-sided alternative is that Spread THieF yields a lower expected loss than the benchmark.

\subsection{Economic evaluation based on BESS arbitrage}
\label{ssec:Econ:Evaluation}

To evaluate the economic benefits of using Spread THieF, we consider a trading strategy that operates a battery energy storage system (BESS) to profit from intraday price differences; for a recent review of BESS trading strategies see \cite{hir:zie:26}. Following \cite{ser:wer:25}, we assume a storage capacity of 1 MWh, equal charging and discharging efficiencies $\eta$, corresponding to a round-trip efficiency of $\eta^2$, and allow a single charge and discharge cycle during the day. The initial state of charge is 0 MWh.

The trading strategy can be formulated as a single optimization problem with a no-trade option. For each day $d$, we either do not operate the battery or choose one hour $h_1$ at which to buy electricity and charge the battery, and a later hour $h_2>h_1$ at which to discharge the battery and sell electricity. The predicted net profit is
\begin{equation}
    \hat{\pi}_{d}(h_1,h_2) = \eta \hat{p}_{d,h_2} - \tfrac{1}{\eta}\hat{p}_{d,h_1} - C,
\label{eqn:pred:profit}
\end{equation}
where $C$ is the round-trip cost of operating a BESS with a capacity of 1 MWh and accounts for investment expenditure, fixed and variable operation and maintenance costs, and battery degradation. Following \cite{lin:zhu:wid:24}, we assume $C=25$ EUR. A similar cost estimate has recently been considered by~\cite{mac:lip:uni:26}.

The optimal decision is therefore
\begin{equation}
    \hat{\pi}_d^* = \max\left\{ 0,\; \max_{1\le i<j\le 24}
        \left(\eta \hat{p}_{d,j} - \tfrac{1}{\eta}\hat{p}_{d,i} - C \right) \right\}.
\label{eqn:opt}
\end{equation}
If $\hat{\pi}_d^*=0$, no trade is executed. Otherwise, the charging and discharging hours are selected as
\begin{equation}
    (h_1,h_2) = \argmax_{1\le i<j\le 24} \left( \eta \hat{p}_{d,j} - \tfrac{1}{\eta}\hat{p}_{d,i} - C \right).
\label{eqn:hours}
\end{equation}
The corresponding realized profit is
\begin{equation}
    \pi_d = \begin{cases}
        \eta p_{d,h_2} - \tfrac{1}{\eta}p_{d,h_1} - C,
        & \text{if } \hat{\pi}_d^*>0,\\[1ex]
        0, & \text{otherwise.}
    \end{cases}
\label{eqn:profit}
\end{equation}
In contrast to \cite{ser:wer:25}, who considered only one efficiency level, $\eta=0.9$, we evaluate three values: $\eta\in\{1,0.95,0.9\}$. The case $\eta=1$ corresponds to a perfectly efficient battery and allows direct comparison with unreconciled spread forecasts. In this case, the efficiency-adjusted margin reduces to the ordinary price spread,
$p_{d,h_2}-p_{d,h_1}$, so spread forecasts alone are sufficient both to select the optimal $(h_1,h_2)$ pair and to determine whether the predicted profit exceeds the operating cost. For $\eta<1$, however, the relevant quantity is
$\eta p_{d,h_2}-\frac{1}{\eta}p_{d,h_1}$, which depends separately on the buying and selling price levels and cannot be recovered from a forecast of the unadjusted spread alone. Consequently, for $\eta\in\{0.95,0.9\}$ we compare the arbitrage profits obtained from Base price forecasts, Block THieF, and Spread THieF, but not from unreconciled spread forecasts.

\subsection{Opportunity cost}
\label{ssec:OC}

We use the \textit{Oracle} (or \textit{Crystal Ball}) strategy as the benchmark for evaluating the economic consequences of forecast errors. Following the common practice of using perfect foresight as an upper benchmark for storage-arbitrage value \citep{mer:oli:dej:23,vee:mul:25}, the Oracle assumes perfect knowledge of day-ahead electricity prices and selects the profit-maximizing BESS decision under the same battery constraints, efficiency, and operating costs as the forecast-based strategies. The Oracle profit on day $d$ is
\begin{equation}
    \pi_d^{\mathrm{Oracle}} = \max\left\{0,\;       \max_{1\le i<j\le24} \left( \eta p_{d,j} -          \tfrac{1}{\eta}p_{d,i} - C \right) \right\}.
\label{eqn:oracle}
\end{equation}
For forecasting method $m$, we define the daily \textit{opportunity cost} as
\begin{equation}
    \mathrm{OC}_{d}^{(m)} = \pi_d^{\mathrm{Oracle}} -    \pi_d^{(m)}.
\label{eqn:oc}
\end{equation}
Thus, $\mathrm{OC}_{d}^{(m)}$ measures the realized profit foregone because the
charging and discharging decision is based on imperfect forecasts rather than perfect foresight. In decision-theoretic terms, this quantity can equivalently be interpreted as \textit{ex-post regret}, i.e., the loss resulting from selecting a suboptimal action under imperfect information \citep{alk:etal:26}. Since the Oracle optimizes over the same feasible set of decisions, including the no-trade option, $\mathrm{OC}_{d}^{(m)}\geq0$.

Because the magnitude of attainable arbitrage profits differs substantially across markets and battery-efficiency levels, we summarize economic performance using the \textit{relative opportunity cost}:
\begin{equation}
    \mathrm{ROC}^{(m)} = 100 \left( 1- \frac{ \sum_{d=1}^{N}
    \pi_d^{(m)}
    }{ \sum_{d=1}^{N}\pi_d^{\mathrm{Oracle}} } \right) \%.
\label{eqn:roc}
\end{equation}
ROC measures the percentage of the maximum attainable profit foregone as a result of forecast-based decisions. This normalization is closely related to the value-of-perfect-information measure used by \citet{pra:bru:mac:25}, who express the revenue loss from forecast-based BESS scheduling relative to perfect-foresight revenue. Lower ROC values therefore indicate better economic performance, with $\mathrm{ROC}=0$ corresponding to perfect foresight. The normalization by Oracle profit also facilitates comparisons across markets and battery-efficiency levels with different arbitrage opportunities.

For comparison with a simple forecasting benchmark, we also consider a \textit{Naive} strategy based on the seasonal naive forecasting method commonly used in EPF \citep{wer:14,lag:mar:sch:wer:21}: $\hat{p}_{d,h}=p_{d-1,h}$ for Tuesday through Friday and $\hat{p}_{d,h}=p_{d-7,h}$ for Saturday through Monday, independently for each hour $h=1,\ldots,24$.

\section{Results}
\label{sec:results}

\begin{table}[tb]
\caption{Forecast evaluation measures for all three models (ARX, NARX and TabPFN) and forecasting frameworks (Base, Block THieF and Spread THieF) over the 5-year test period in the German and Spanish markets; see Sec.~\ref{ssec:Stat:Errors} for error metric definitions. The coloring is applied jointly for all Skill Scores. 
For each forecasting model, Spread THieF significantly outperforms both Base and Block THieF at the 1\% level under both absolute-error and squared-error loss, according to the CPA test; see Sec.~\ref{ssec:Stat:CPA} for details.\\[-6pt]}
\label{tab:DE:ES}
\resizebox{\textwidth}{!}{%
\begin{tabular}{ccccccrrrr}
    \toprule
\multicolumn{2}{l}{} & {RMSE} & {dRMSE} & {MAE} & {dMAE} &  $\text{SS}_{\text{RMSE}}$ & $\text{SS}_{\text{dRMSE}}$  &    $\text{SS}_{\text{MAE}}$ & $\text{SS}_{\text{dMAE}}$ \\
\multicolumn{10}{c}{\textit{Germany}} \\
\midrule

                         & Base                 & 35.92                    & 28.19                     & 23.65                   & 19.18                    & \cellcolor[HTML]{FCFCFF}---  & \cellcolor[HTML]{FCFCFF}---  & \cellcolor[HTML]{FCFCFF}---  & \cellcolor[HTML]{FCFCFF}---  \\
                         & Block THieF          & 34.89                    & 27.22                     & 22.97                   & 18.65                    & \cellcolor[HTML]{E7F4ED}2.9\%  & \cellcolor[HTML]{E2F2E9}3.4\%  & \cellcolor[HTML]{E7F4ED}2.9\%  & \cellcolor[HTML]{E7F4ED}2.8\%  \\
\multirow{-3}{*}{ARX}    & Spread THieF         & 30.49                    & 22.71                     & 19.77                   & 15.39                    & \cellcolor[HTML]{89CE9C}15.1\% & \cellcolor[HTML]{68C07F}19.4\% & \cellcolor[HTML]{7FCA93}16.4\% & \cellcolor[HTML]{65BF7D}19.7\% \\
\midrule
                         & Base                 & 29.58                    & 22.06                     & 18.21                   & 14.03                    & \cellcolor[HTML]{FCFCFF}---  & \cellcolor[HTML]{FCFCFF}---  & \cellcolor[HTML]{FCFCFF}---  & \cellcolor[HTML]{FCFCFF}---  \\
                         & Block THieF          & 28.10                    & 20.26                     & 17.29                   & 12.94                    & \cellcolor[HTML]{D6EDDF}5.0\%  & \cellcolor[HTML]{BEE3CA}8.2\%  & \cellcolor[HTML]{D6EDDE}5.0\%  & \cellcolor[HTML]{C1E4CC}7.8\%  \\
\multirow{-3}{*}{NARX}   & Spread THieF         & 26.45                    & 19.39                     & 16.02                   & 12.10                    & \cellcolor[HTML]{ACDCBA}10.6\% & \cellcolor[HTML]{A0D7B0}12.1\% & \cellcolor[HTML]{A1D7B0}12.0\% & \cellcolor[HTML]{93D2A4}13.8\% \\
\midrule
                         & Base                 & 27.30                    & 20.19                     & 15.49                   & 11.80                    & \cellcolor[HTML]{FCFCFF}---  & \cellcolor[HTML]{FCFCFF}---  & \cellcolor[HTML]{FCFCFF}---  & \cellcolor[HTML]{FCFCFF}---  \\
                         & Block THieF          & 27.59                    & 20.59                     & 15.79                   & 12.03                    & \cellcolor[HTML]{FBECEF}$-$1.1\% & \cellcolor[HTML]{FBDEE1}$-$2.0\% & \cellcolor[HTML]{FBDFE2}$-$1.9\% & \cellcolor[HTML]{FBDFE2}$-$1.9\% \\
\multirow{-3}{*}{TabPFN} & Spread THieF         & 25.69                    & 18.71                     & 15.22                   & 11.16                    & \cellcolor[HTML]{CFEAD9}5.9\%  & \cellcolor[HTML]{C4E6CF}7.3\%  & \cellcolor[HTML]{EFF7F4}1.8\%  & \cellcolor[HTML]{D3ECDC}5.4\% \\
	\midrule
    \multicolumn{10}{c}{\textit{Spain}} \\
    \midrule
                         & Base                 & 23.09                    & 17.84                     & 16.31                   & 12.64                    & \cellcolor[HTML]{FCFCFF}---  & \cellcolor[HTML]{FCFCFF}---  & \cellcolor[HTML]{FCFCFF}---  & \cellcolor[HTML]{FCFCFF}---  \\
                         & Block THieF          & 22.72                    & 17.52                     & 15.92                   & 12.36                    & \cellcolor[HTML]{F0F8F5}1.6\%  & \cellcolor[HTML]{EFF7F4}1.8\%  & \cellcolor[HTML]{EAF5F0}2.4\%  & \cellcolor[HTML]{ECF6F1}2.1\% \\
\multirow{-3}{*}{ARX}    & Spread THieF         & 21.09                    & 16.42                     & 14.56                   & 11.23                    & \cellcolor[HTML]{BAE2C6}8.6\%  & \cellcolor[HTML]{C0E4CB}7.9\%  & \cellcolor[HTML]{ABDBB9}10.7\% & \cellcolor[HTML]{A8DAB6}11.1\% \\
\midrule
                         & Base                 & 22.25                    & 17.22                     & 15.53                   & 12.09                    & \cellcolor[HTML]{FCFCFF}---  & \cellcolor[HTML]{FCFCFF}---  & \cellcolor[HTML]{FCFCFF}---  & \cellcolor[HTML]{FCFCFF}---  \\
                         & Block THieF          & 21.53                    & 16.36                     & 15.00                   & 11.41                    & \cellcolor[HTML]{E4F2EA}3.2\%  & \cellcolor[HTML]{D6EDDF}5.0\%  & \cellcolor[HTML]{E3F2E9}3.4\%  & \cellcolor[HTML]{D2EBDA}5.6\%  \\
\multirow{-3}{*}{NARX}   & Spread THieF         & 20.30                    & 15.90                     & 14.18                   & 10.98                    & \cellcolor[HTML]{B9E1C6}8.8\%  & \cellcolor[HTML]{C2E5CD}7.7\%  & \cellcolor[HTML]{BAE2C6}8.7\%  & \cellcolor[HTML]{B6E0C3}9.2\%  \\
\midrule
                         & Base                 & 19.86                    & 15.30                     & 13.24                   & 10.13                    & \cellcolor[HTML]{FCFCFF}---  & \cellcolor[HTML]{FCFCFF}---  & \cellcolor[HTML]{FCFCFF}---  & \cellcolor[HTML]{FCFCFF}---  \\
                         & Block THieF          & 19.90                    & 15.31                     & 13.35                   & 10.18                    & \cellcolor[HTML]{FBF9FC}$-$0.2\% & \cellcolor[HTML]{FBFBFE}$-$0.1\% & \cellcolor[HTML]{FBEFF2}$-$0.8\% & \cellcolor[HTML]{FBF5F8}$-$0.5\% \\
\multirow{-3}{*}{TabPFN} & Spread THieF         & 19.12                    & 14.67                     & 12.85                   & 9.54                     & \cellcolor[HTML]{E0F1E7}3.7\%  & \cellcolor[HTML]{DDF0E4}4.1\%  & \cellcolor[HTML]{E6F3EC}2.9\%  & \cellcolor[HTML]{D0EAD9}5.8\% \\
\bottomrule    
	\end{tabular}
}
\end{table}

\subsection{Stealing accuracy with THieF}
\label{ssec:results:stat}

Table~\ref{tab:DE:ES} reports the forecast accuracy of the standard hourly forecasts (Base) and the two hierarchical approaches (Block THieF and Spread THieF) for the German and Spanish markets, respectively. In Germany, Spread THieF improves all four error measures for all three forecasting models.  Relative to the corresponding Base forecasts, the gains range from 1.8\% to 16.4\% for the hourly price measures and from 5.4\% to 19.7\% for the daily-average-price measures. The largest improvements are obtained for ARX, the simplest and least accurate base model. This is in agreement with the literature, where THieF gains correlate with the degree of the misspecification of the forecasting model \citep[e.g., see][]{ath:hyn:kou:pet:17}. Importantly, Spread THieF also improves the already competitive TabPFN forecasts, reducing hourly and daily RMSE by 5.9\% and 7.3\%, respectively. In contrast, Block THieF provides positive gains for ARX and NARX but slightly decreases the accuracy of TabPFN in terms of all four error measures.

The results for the Spanish market lead to a similar conclusion. Spread THieF improves the hourly price measures by 2.9\%-10.7\% and the daily average price measures by 4.1\%-11.1\% relative to Base forecasts. Again, the largest gains are obtained for ARX, while the improvements for TabPFN are smaller but consistently positive. In both markets, TabPFN remains the most accurate forecasting architecture and ARX the least accurate. Nevertheless, in Spain ARX combined with Spread THieF outperforms Base NARX under all four accuracy measures and also outperforms NARX with Block THieF under all measures except dRMSE. This shows that incorporating direct forecasts of intraday price differences through reconciliation can compensate for part of the performance gap between simpler and more complex base forecasting models.

The statistical superiority of Spread THieF is remarkably robust. For every forecasting model in both Germany and Spain, the Conditional Predictive Ability (CPA) tests reject equal predictive ability in favor of Spread THieF at the 1\% significance level, both against Base forecasts and against Block THieF, under absolute-error and squared-error loss. The contrast with TabPFN is particularly informative: while conventional Block THieF slightly worsens the highly accurate TabPFN Base forecasts in both markets, Spread THieF improves them under every evaluation measure. This suggests that direct spread forecasts contribute information that is not provided by temporal aggregation of price levels alone.

\begin{figure}[tbp]
    \centering
    \includegraphics[width=\linewidth]{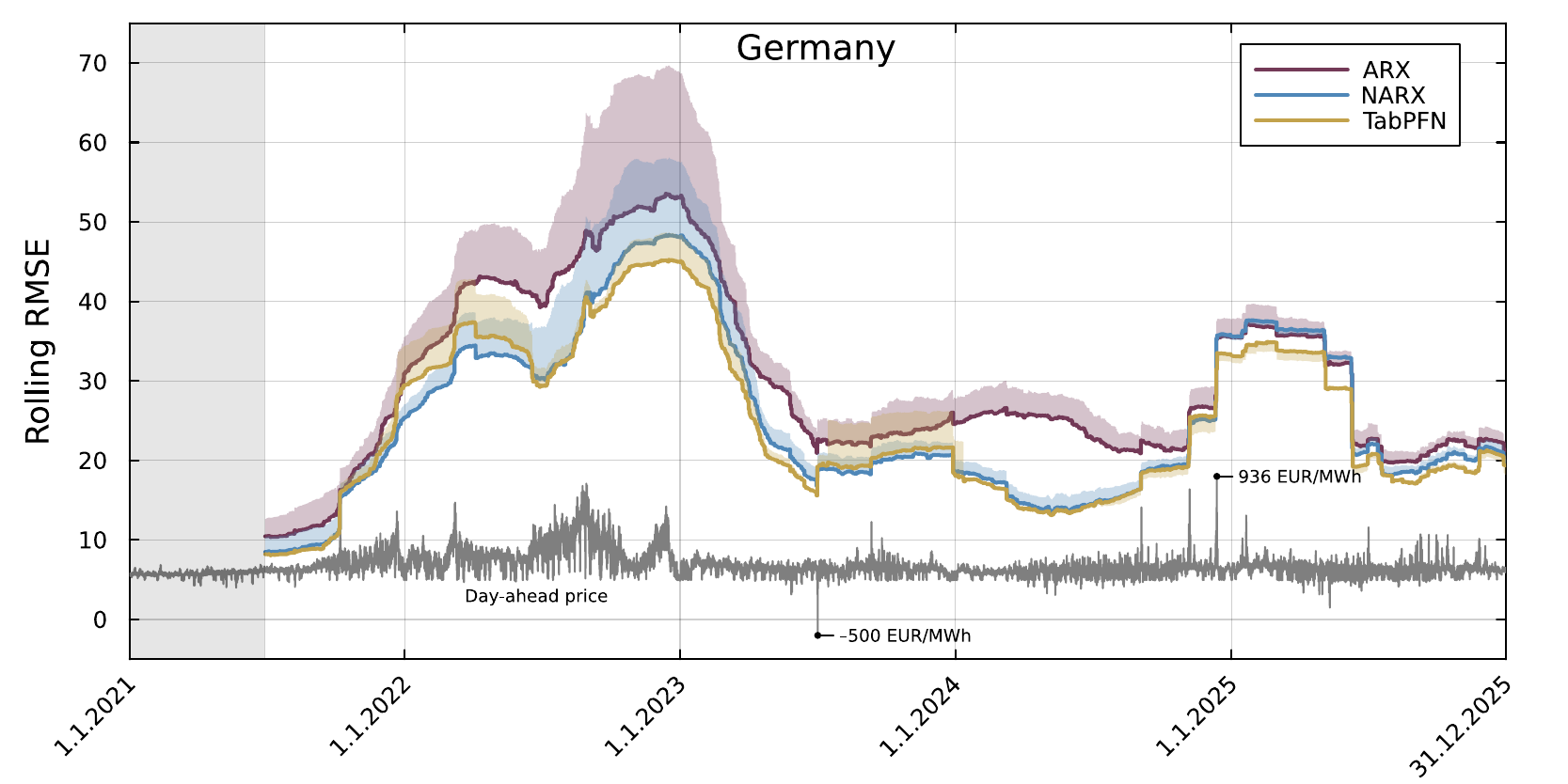}\vspace{-1.2cm}
    \includegraphics[width=\linewidth]{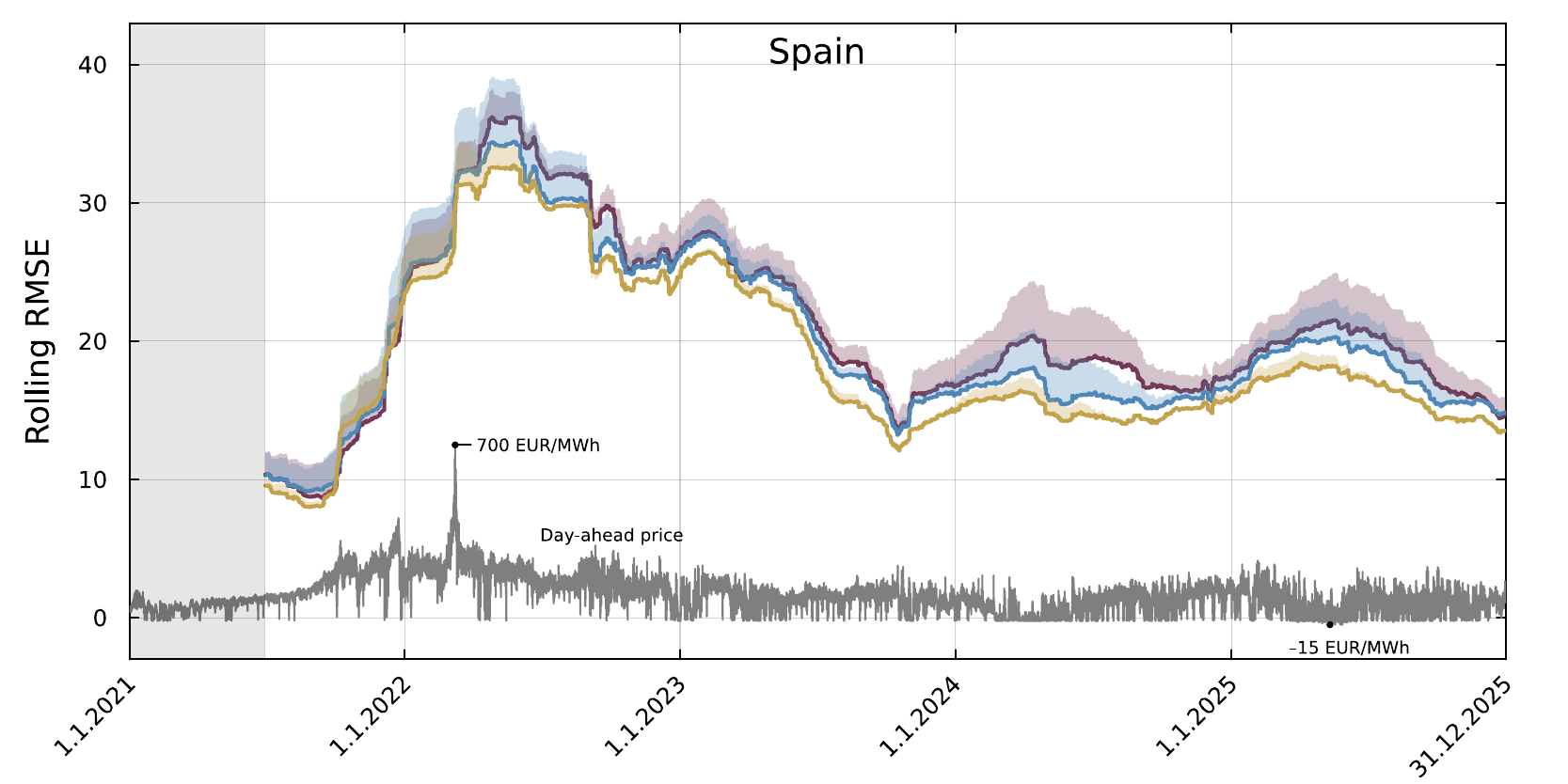}
    \caption{Rolling RMSE over a 180-day window in the German (\textit{top}) and Spanish (\textit{bottom}) markets. The solid lines mark the RMSE of the Spread THieF approach for the three models. The lighter shaded areas of the same color extend from the Spread THieF value to the corresponding Base value, so that their width represents the improvement obtained from reconciliation. Values for 29.06.2021 are the RMSE for 1.01-29.06.2021, i.e., the light gray shaded area on the left, values for 30.06.2021 are the RMSE for 2.01-30.06.2021, etc. The gray line at the bottom of each plot shows the day-ahead price series, rescaled for visualization and with extreme prices annotated.}
    \label{fig:rmse}
\end{figure}

Figure~\ref{fig:rmse} examines whether these improvements are stable over time by reporting RMSE over rolling 180-day windows. Forecast difficulty varies substantially throughout the test period, particularly around the large price movements of 2021-2022. Nevertheless, the gains from Spread THieF are not confined to a single episode and are observed across markedly different market regimes. The improvements are generally largest for ARX, consistent with the full-sample results in Table~\ref{tab:DE:ES}, while TabPFN remains the most accurate model over most of the sample. 

\begin{table}[tb]
\caption{Total profits (EUR) from the BESS trading strategy for all three models (ARX, NARX, and TabPFN) and the considered forecasting frameworks over the 5-year test period in the German and Spanish markets. Note that for efficiency $\eta<1$ the decisions cannot be recovered from forecasts of the unadjusted spreads alone.
The coloring is applied separately for each market and efficiency. For comparison, we report the profits for the Oracle strategy, which assumes perfect foresight, and the strategy based on Naive price forecasts; see Sec.~\ref{ssec:Econ:Evaluation} and \ref{ssec:OC} for details.\\[-6pt]}
\label{tab:strategy}
\centering
\small{
\begin{tabular}{lrrrrrr}
    \toprule
    & ARX                             & NARX                            & TabPFN                          & ARX                            & NARX                           & TabPFN                         \\
    \multicolumn{1}{l}{} & \multicolumn{3}{c}{\textit{Germany}} & \multicolumn{3}{c}{\textit{Spain}}    \\
\midrule
\multicolumn{1}{l}{} & \multicolumn{6}{c}{$\eta = 1$} \\[3pt]
Base (hours)         & \cellcolor[HTML]{FCFCFF}150,281 & \cellcolor[HTML]{FCFCFF}148,743 & \cellcolor[HTML]{EEF7F3}151,350 & \cellcolor[HTML]{DFF1E6}80,118 & \cellcolor[HTML]{FCFCFF}77,552 & \cellcolor[HTML]{CAE8D4}80,410 \\
Base (spreads)         & \cellcolor[HTML]{FCFCFF}149,442 & \cellcolor[HTML]{FCFCFF}149,662 & \cellcolor[HTML]{FCFCFF}147,815 & \cellcolor[HTML]{BCE2C8}80,616 & \cellcolor[HTML]{F6FAFA}79,796 & \cellcolor[HTML]{FCFCFF}78,055 \\
Block THieF               & \cellcolor[HTML]{F4F9F9}151,211 & \cellcolor[HTML]{FCFCFF}149,448 & \cellcolor[HTML]{DFF1E6}151,662 & \cellcolor[HTML]{D4ECDD}80,272 & \cellcolor[HTML]{FCFCFF}77,975 & \cellcolor[HTML]{DEF0E5}80,140 \\
Spread THieF           & \cellcolor[HTML]{92D2A4}153,274 & \cellcolor[HTML]{63BE7B}154,265 & \cellcolor[HTML]{89CD9B}153,485 & \cellcolor[HTML]{6EC385}81,704 & \cellcolor[HTML]{63BE7B}81,859 & \cellcolor[HTML]{A1D7B0}81,000 \\
Oracle & & 166,424 & & & 95,336 & \\
Naive  & & 132,684 & & & 71,072 & \\
\midrule
\multicolumn{1}{l}{} & \multicolumn{6}{c}{$\eta = 0.95$} \\[3pt]
Base (hours)         & \cellcolor[HTML]{FCFCFF}127,198 & \cellcolor[HTML]{FCFCFF}126,025 & \cellcolor[HTML]{FCFCFF}128,589 & \cellcolor[HTML]{FCFCFF}61,892 & \cellcolor[HTML]{FCFCFF}59,770 & \cellcolor[HTML]{CEEAD8}62,536 \\
Block THieF          & \cellcolor[HTML]{FCFCFF}128,072 & \cellcolor[HTML]{FCFCFF}126,691 & \cellcolor[HTML]{FAFBFD}128,788 & \cellcolor[HTML]{EAF5EF}62,152 & \cellcolor[HTML]{FCFCFF}59,979 & \cellcolor[HTML]{E3F2E9}62,252 \\
Spread THieF         & \cellcolor[HTML]{ABDBB9}130,181 & \cellcolor[HTML]{63BE7B}131,449 & \cellcolor[HTML]{91D1A3}130,645 & \cellcolor[HTML]{75C68B}63,770 & \cellcolor[HTML]{63BE7B}64,019 & \cellcolor[HTML]{A9DBB8}63,048 \\
Oracle & & 143,474 & & & 77,558 & \\
Naive  & & 109,621 & & & 52,758 & \\
\midrule
\multicolumn{1}{l}{} & \multicolumn{6}{c}{$\eta = 0.9$} \\[3pt]
Base (hours)         & \cellcolor[HTML]{FCFCFF}104,395 & \cellcolor[HTML]{FCFCFF}103,251 & \cellcolor[HTML]{F8FBFC}106,047 & \cellcolor[HTML]{FCFCFF}45,307 & \cellcolor[HTML]{FCFCFF}43,261 & \cellcolor[HTML]{B0DEBE}46,618 \\
Block THieF          & \cellcolor[HTML]{FCFCFF}105,217 & \cellcolor[HTML]{FCFCFF}103,926 & \cellcolor[HTML]{EAF5F0}106,292 & \cellcolor[HTML]{FCFCFF}45,509 & \cellcolor[HTML]{FCFCFF}43,514 & \cellcolor[HTML]{C2E5CD}46,365 \\
Spread THieF         & \cellcolor[HTML]{ABDBB9}107,420 & \cellcolor[HTML]{63BE7B}108,681 & \cellcolor[HTML]{92D1A4}107,856 & \cellcolor[HTML]{7DC991}47,377 & \cellcolor[HTML]{63BE7B}47,746 & \cellcolor[HTML]{83CB97}47,282 \\
Oracle & & 120,827 & & & 61,158 & \\
Naive  & & 86,317 & & & 35,495 & \\
\bottomrule
\end{tabular}
}
\end{table}

\subsection{Stealing profits with THieF}
\label{ssec:results:econ}

The economic results in Table~\ref{tab:strategy} reinforce the findings in terms of statistical error metrics. For every model, market, and battery-efficiency level, Spread THieF yields a higher realized profit and a lower relative opportunity cost than the corresponding Base and Block THieF forecasts. For $\eta=1$, it also outperforms the unreconciled spread forecasts. Moreover, even the least profitable Spread THieF specification outperforms the best non-Spread alternative for every market and efficiency level $\eta\in\{1,0.95,0.9\}$. Relative to Block THieF based on the same forecasting model, the profit gains range from approximately 1.1\% to 9.7\%.

Interestingly, the economic ranking of the forecasting architectures differs from their statistical ranking. Although TabPFN provides the most accurate price forecasts, NARX combined with Spread THieF yields the highest profit, and hence the lowest opportunity cost, in all six market-efficiency combinations. This confirms that improvements in conventional forecast error measures need not translate one-to-one into improvements in trading decisions. 

\begin{figure}[tbp]
    \centering
    \includegraphics[width=\linewidth]{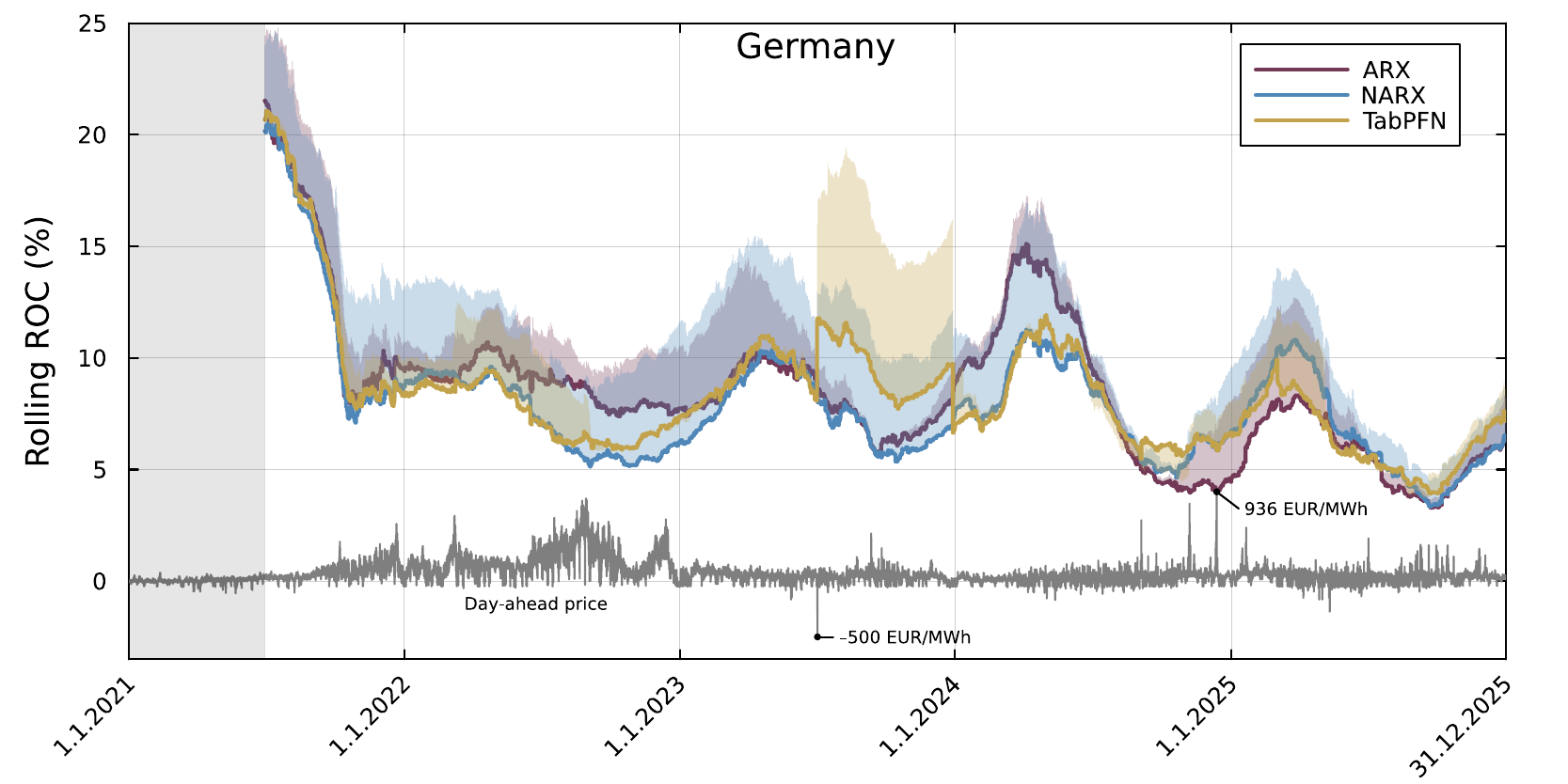}\vspace{-1.2cm}
    \includegraphics[width=\linewidth]{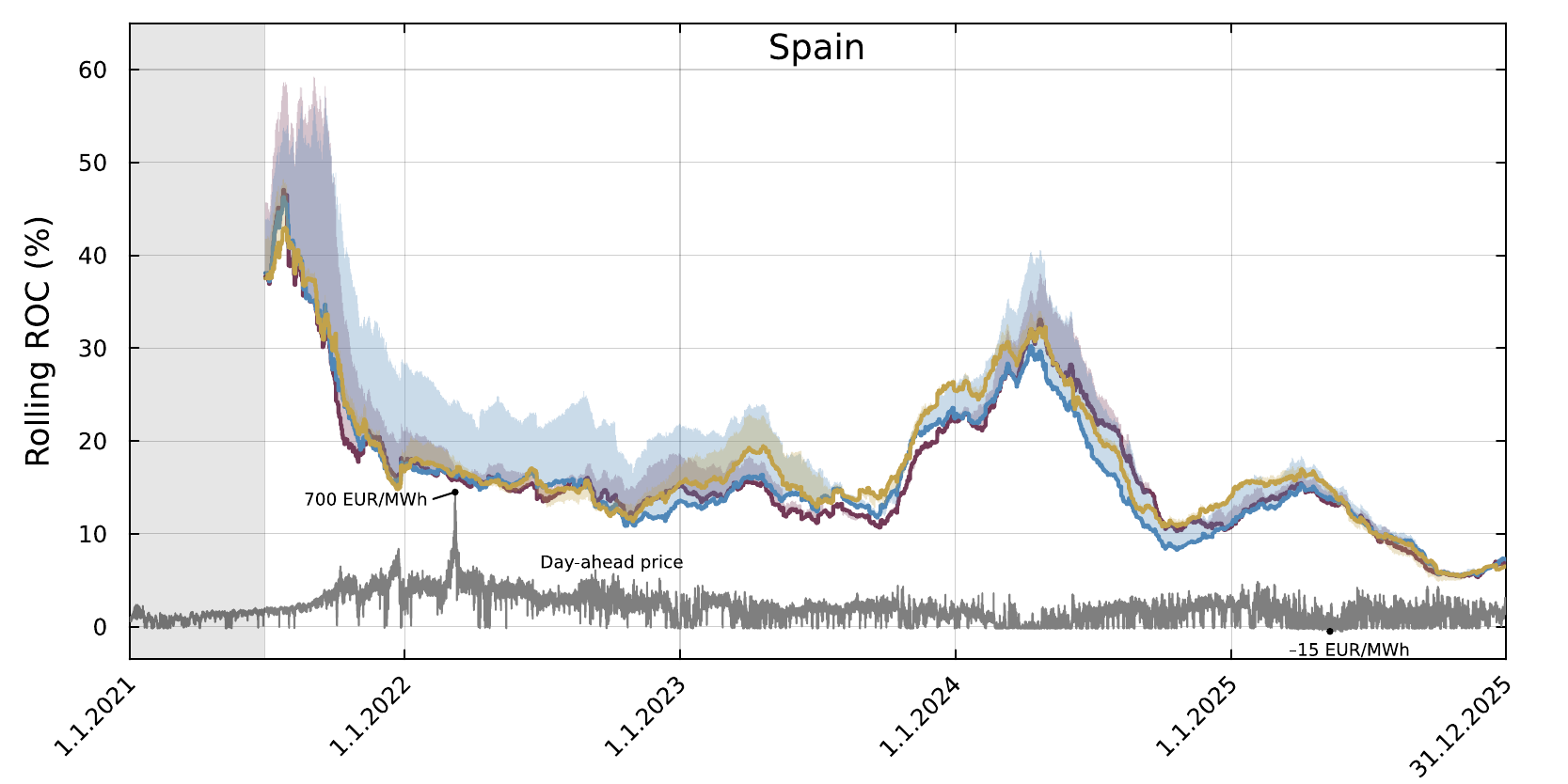}
    \caption{Rolling Relative Opportunity Cost (ROC) over a 180-day window in the German (\textit{top}) and Spanish (\textit{bottom}) markets, and battery efficiency $\eta=1$. The solid lines mark the ROC of the Spread THieF approach for the three models. The lighter shaded areas of the same color extend from the Spread THieF value to the corresponding Base value, so that their width represents the improvement obtained from reconciliation. Values for 29.06.2021 are the ROC scores for 1.01-29.06.2021, i.e., the light gray shaded area on the left, values for 30.06.2021 are the ROC scores for 2.01-30.06.2021, etc. The gray line at the bottom of each plot shows the day-ahead price series, rescaled for visualization and with extreme prices annotated. The elevated rolling ROC for TabPFN in the German market in the second half of 2023 is largely driven by a single trading decision, see Fig.~\ref{fig:trade}.}
    \label{fig:oc}
\end{figure}

Expressed in terms of relative opportunity cost (ROC), Spread THieF loses approximately 7.3\%-11.1\% of the Oracle profit in Germany and 14.1\%-22.7\% in Spain, depending on the forecasting model and battery efficiency. 
The opportunity cost gap between the two markets is much larger than suggested by their statistical forecast errors. This further illustrates that the economic consequences of forecast errors depend not only on their magnitude but also on whether they alter the economically relevant ordering of prices within the day.
Relative to Block THieF, Spread THieF reduces ROC by approximately 0.9-6.9 percentage points. 
Figure~\ref{fig:oc} shows the corresponding rolling ROC for $\eta=1$ over 180-day windows. The economic performance varies considerably over time, and the evolution of rolling ROC does not simply replicate that of rolling RMSE. This further highlights the distinction between statistical forecast accuracy and the economic value of the resulting trading decisions.

\begin{figure}
    \centering
    \includegraphics[width=\linewidth]{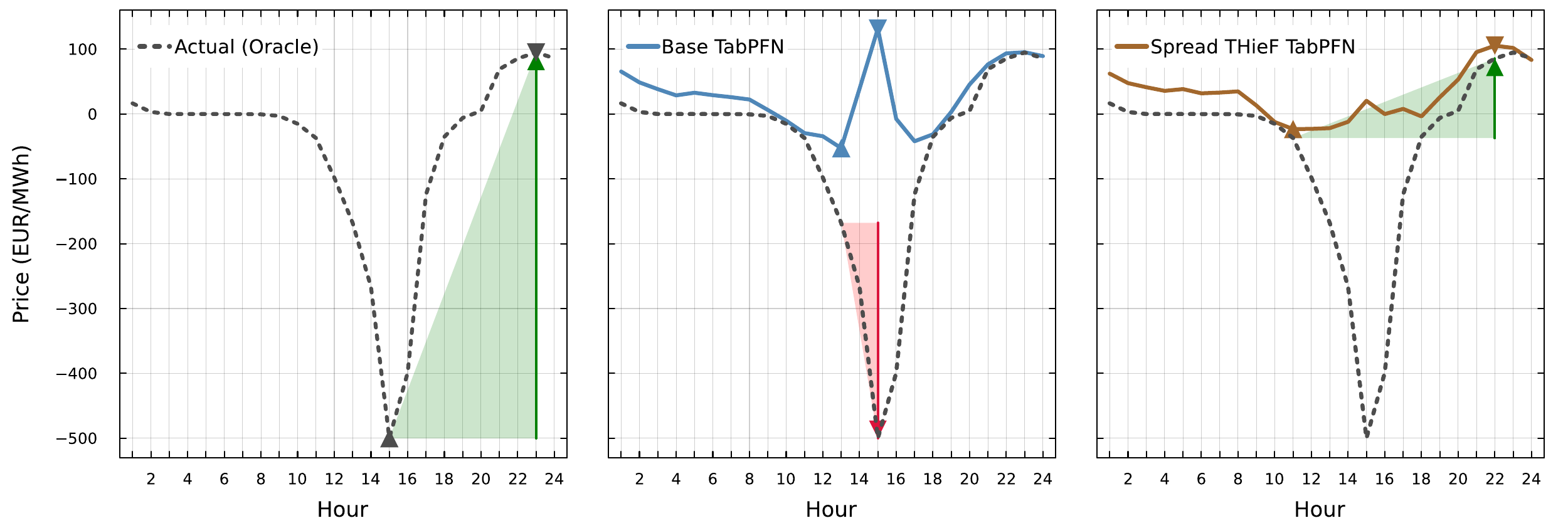}
    \caption{Example of trading decisions in the German day-ahead market on Sunday 2 July 2023, when the price reached its minimum over the test period, falling to $-500.00$ EUR/MWh in hour 15. Dashed lines represent realized prices, which are used by the perfect-foresight Oracle benchmark, while solid lines show TabPFN-2 predictions. Colored triangles indicate trading decisions, with green denoting a profitable trade and red a loss. Based on the unreconciled hourly forecasts, TabPFN selects hour 15 as the selling hour, resulting in a single-day loss exceeding 300 EUR for the 1 MWh BESS. In contrast, Spread THieF adjusts the predicted daily price profile sufficiently to avoid this costly trading decision and instead yields a positive profit.}
    \label{fig:trade}
\end{figure}

The elevated rolling ROC for TabPFN in the German market in the second half of 2023 is largely driven by a single trading decision. On Sunday 2 July 2023, high renewable generation combined with low demand led to strongly negative day-ahead prices, with a daily average of $-53.9$ EUR/MWh and a minimum of $-500$ EUR/MWh in hour 15. Figure~\ref{fig:trade} illustrates the corresponding trading decisions based on Base TabPFN, Spread THieF TabPFN, and the Oracle benchmark. The Base TabPFN forecast incorrectly identifies hour 15 as the optimal selling hour, resulting in a loss exceeding EUR~300 for the 1~MWh BESS. In contrast, Spread THieF modifies the relative shape of the predicted daily price profile sufficiently to avoid this costly decision and instead yields a positive profit. This example illustrates how improving the relative relationships between hourly prices can have substantial economic value even when the change in aggregate error measures is comparatively modest.

\section{Conclusions}
\label{sec:conclusions}

In day-ahead electricity markets, the economic value of a price forecast often depends not only on the accuracy of individual hourly prices but also on their relative differences within the day. Motivated by this observation, we extend temporal hierarchy forecasting (THieF) by combining forecasts of hourly prices with direct forecasts of all pairwise intraday price spreads. The resulting Spread THieF framework exploits information on both absolute price levels and relative intraday price movements, while forecast reconciliation ensures coherence between them.

Using five years of out-of-sample data from the German and Spanish markets, we find that Spread THieF consistently improves forecast accuracy across three markedly different forecasting architectures: a linear ARX model, a shallow neural network, and the TabPFN-2 foundation model. Relative to the corresponding Base forecasts, the gains range from approximately 2\% to 20\% across markets and error measures. Importantly, Spread THieF also improves the highly accurate TabPFN-2 forecasts, whereas conventional Block THieF slightly decreases their accuracy in both markets. Moreover, for every forecasting model and in both markets, Spread THieF significantly outperforms both Base and Block THieF at the 1\% level under both absolute-error and squared-error loss.

The statistical improvements translate into economically meaningful gains in the stylized BESS arbitrage application. Within each forecasting architecture, Spread THieF yields the lowest opportunity cost in both markets and under all three battery-efficiency assumptions considered. Relative to Block THieF, it reduces relative opportunity cost by approximately 0.9-6.9 percentage points, while the corresponding profit gains range from approximately 1.1\% to 9.7\%. For every market and battery-efficiency level, even the least profitable Spread THieF specification outperforms the best competing forecasting approach. The gains also persist across markedly different market regimes, while the rolling analysis shows that statistical forecast accuracy and economic performance can evolve quite differently over time.
At the same time, the economic and statistical rankings of the forecasting architectures differ: although TabPFN is the most accurate model, NARX combined with Spread THieF yields the highest arbitrage profit in every market-efficiency combination. This highlights the importance of evaluating electricity price forecasts not only statistically but also in terms of the decisions they support.

The BESS exercise is deliberately stylized: we assume price-taking behavior, a single daily charge-discharge cycle, and a fixed round-trip cost. Future research could examine whether the economic gains persist under more realistic operational constraints and alternative trading strategies. On the forecasting side, a natural extension is to incorporate battery efficiency directly into the reconciliation structure through efficiency-adjusted spreads, rather than introducing it only at the decision stage. Another promising direction is to combine spread information with temporally aggregated price differences, thereby integrating the relative-price information considered here with the multiscale structure of conventional THieF.

\section*{Acknowledgments}

The study was partially supported by the National Science Center (NCN, Poland) through grant no.\ 2025/57/B/HS4/02413.

\bibliographystyle{elsarticle-harv} 
\begin{footnotesize}
\setlength{\bibsep}{3pt}
\bibliography{epf}
\end{footnotesize}

\end{document}